\documentclass[twocolumn]{aastex61}
\pdfoutput=1

\usepackage{amsmath}
\usepackage{comment}
\usepackage{savesym}
\savesymbol{tablenum}
\usepackage{siunitx}
\restoresymbol{SIX}{tablenum}

\newcommand{\refsec}[1]{\mbox{\S\ \ref{sec:#1}}}

\newcommand{\ctbd}[1]{}

\newcommand\poet{\textsc{Poet}}
\defcitealias{Penev_et_al_16}{P16} 
\newcommand{\Ntotalattempted}{188}

\newcommand{\Ntwosidedtotal}{35}
\newcommand{\Nlowerlim}{40}

\begin{document}

\title{Empirical Tidal Dissipation in Exoplanet Hosts From Tidal Spin--Up}

\shorttitle{Measuring $Q_\star'$ From Tidal Spin-Up}

\author[0000-0003-4464-1371]{Kaloyan Penev}
\affiliation{
    Department of Physics,
    The University of Texas at Dallas,
    800 West Campbell Road,
    Richardson, TX 75080-3021
    USA
}

\author[0000-0002-0514-5538]{L. G. Bouma}
\affiliation{
    Department of Astrophysical Sciences,
    Princeton University,
    4 Ivy Ln.,
    Princeton, NJ 08544,
    USA
}

\author[0000-0002-4265-047X]{Joshua N.\ Winn}
\affiliation{
    Department of Astrophysical Sciences,
    Princeton University,
    4 Ivy Ln.,
    Princeton, NJ 08544,
    USA
}

\author[0000-0001-8732-6166]{Joel D. Hartman}
\affiliation{
    Department of Astrophysical Sciences,
    Princeton University,
    4 Ivy Ln.,
    Princeton, NJ 08544,
    USA
}

\correspondingauthor{Kaloyan Penev}
\email{kaloyan.penev@utdallas.edu}


\begin{abstract}
    Stars with hot Jupiters tend to be rotating faster than other stars of the same
age and mass. This trend has been attributed to tidal interactions between the
star and planet.  A constraint on the dissipation parameter $Q_\star'$ follows
from the assumption that tides have managed to spin up the star to the observed
rate within the age of the system.  This technique was applied previously to
HATS-18 and WASP-19.  Here we analyze the sample of all \Ntotalattempted\ known
hot Jupiters with an orbital period $<$3.5~days and a ``cool'' host star
($T_{\rm eff}<6100$\,K).  We find evidence that the tidal dissipation parameter
($Q_\star'$) increases sharply with forcing frequency, from $10^5$ at
0.5~day$^{-1}$ to $10^7$ at 2~day$^{-1}$. This helps to resolve a number of
apparent discrepancies between studies of tidal dissipation in binary stars, hot
Jupiters, and warm Jupiters.  It may also allow for a hot Jupiter to damp the
obliquity of its host star prior to being destroyed by tidal decay.

\end{abstract}

\section{Introduction}

Tidal forces allow astronomical bodies to exchange energy and angular
momentum. The friction associated with tidally-induced fluid flow leads to
long-term energy dissipation, with profound consequences throughout all of
astrophysics.  Even when restricting our attention to low-mass stars, there
are numerous situations in which tidal dissipation plays a pivotal role:
\begin{itemize}
    \item A classic question in exoplanetary science is how hot Jupiters
        (HJs) attain their tight orbits. One proposed answer is that
        wide-orbiting giant planets can be thrown into high-eccentricity
        orbits with small periastron distances, which are then shrunk and
        circularized by tidal dissipation in the planet and the star
        \citep{Rasio_Ford_96}.  Another mechanism for HJ placement invokes
        interactions with the protoplanetary disk, causing the planet's orbit
        to spiral inward \citep{Lin_et_al_96}.  In this scenario the planet
        arrives very early, when the star is still young and large, and tidal
        dissipation threatens to cause the planet to spiral further inward
        and become engulfed.  \citet{Nelson_et_al_2017} recently argued that
        the observed distribution of orbital distances suggests that both
        mechanisms might operate.
    \item Tidally-induced orbital decay has been implicated in the apparent
        lack of short-period giant planets in globular clusters
        \citep{Debes_Jackson_10} and around sub-giant stars
        \citep{Schlaufman_Winn_13}. There have also been reports of direct
        detection of period shrinkage for hot Jupiters \citep[{\it
        e.g.},][]{Maciejewski_et_al_16, Patra_et_al_17}, although none have
        been confirmed beyond doubt.
    \item Many HJs have been found to have orbits that are misaligned with
        the host star's equatorial plane. The patterns observed within the
        collection of stellar-obliquity measurements have led to the
        hypothesis that tidal dissipation has re-aligned many systems that
        were formerly misaligned \citep[c.f.][]{Winn_et_al_10,
        Valsecchi_Rasio_14}.
    \item The mechanisms responsible for tidal dissipation must also operate
        on stellar pulsations and oscillations. In particular,
        \citet{Gonczi_81} showed that the red limit of the Cepheid
        instability strip is very sensitive to the dissipation efficiency,
        and \citet{Goldreich_Kumar_88} and \citet{Goldreich_Murray_Kumar_94}
        presented a theory for the amplitudes of solar $p$-modes in which the
        results are very sensitive to the dissipation model.
\end{itemize}

Clearly, a better understanding of the dissipative processes in stars would have
broad implications. The endpoint of tidal evolution can often be predicted based
on simple considerations of energy and angular momentum, but the rate of
evolution cannot yet be predicted from first principles, owing to the complexity
and uncertainty in the processes that convert large-scale motions into heat. The
rate is traditionally parameterized by the dimensionless quality factor
$Q_\star' \equiv Q_\star/k_2$ \citep{Goldreich_Soter_1966}, where $Q_\star$ is
the inverse of the phase lag between the tidal potential and the tidal bulge,
and $k_2$ is the tidal Love number. There are many theoretical models for tidal
dissipation in circulation, giving contradictory predictions for this parameter
\citep[cf.][]{Zahn_75, Zahn_89, Goldreich_Keeley_77, Goodman_Dickson_98,
Ogilvie_Lin_04, Penev_Barranco_Sasselov_09, Essick_Weinberg_16}.

This work is concerned with tidal interactions between hot Jupiters and their
host stars.  The very shortest-period giant planets ($P \lesssim$~few days)
usually have circular orbits, as expected theoretically due to tidal dissipation
within the planet.  This represents a state of minimum orbital energy at fixed
angular momentum.  After reaching this stage, further changes to the orbit must
involve the transfer of orbital angular momentum to the rotation of either the
star or the planet. Because the planet's rotational angular momentum is many
orders of magnitude smaller than that of the orbit, dissipation within the
planet cannot possibly cause significant orbital evolution after the orbit is
circularized. Stars, on the other hand, have much larger masses and sizes and
can exchange large amounts of angular momentum with the orbit. Thus, for the
large number of HJs with circular orbits, only the tidal dissipation within the
star is important. Dissipation within the star leads to changes in the stellar
spin and the orbital period. For the common case in which the stellar rotation
period is longer than the orbital period, tides cause the star to spin faster
and the orbit to shrink.

\citet{Pont_09} and \citet{Husnoo_et_al_12} presented evidence that the stars
hosting some of the shortest-period planets have indeed been tidally spun up.
Since those studies, many more systems showing excess stellar rotation have been
found. \citet{Penev_et_al_16} (\citetalias{Penev_et_al_16}) examined two systems
with unusually short periods. Under the assumption that tides have spun up the
stars within their estimated ages, they found the allowed range for $\log
Q_\star'$ to be 7.2--7.4 for HATS-18 (\citetalias{Penev_et_al_16}), and 6.5--6.9
for WASP-19 \citep{Hebb_et_al_10}.  These and most other previous studies have
assumed that the $Q_\star'$ parameter is independent of the frequency of the
tidal forcing over the range of interest, even though there is no physical
reason to expect this to be the case.  Indeed there are strong theoretical
reasons to expect $Q_\star'$ to be a sensitive function of frequency \citep[see,
e.g.,][]{Goldreich_63,Ogilvie_14}. 

In this study, we improve on the sophistication of the
\citetalias{Penev_et_al_16} analysis, and apply it to 75 systems rather than
two. Section~\ref{sec:method} describes our method for modeling the combined
orbital and stellar spin evolution and deriving constraints on $Q_\star'$.
Section~\ref{sec:results} presents the results.  Section~\ref{sec:discussion}
places the results within the context of other available constraints on
$Q_\star'$, and discusses the implications for HJ systems, including the
possibility of tidal realignment of the stellar obliquity in HJ systems.

\section{Methods}
\label{sec:method}

\begin{figure}[t!]
	\begin{center}
		\includegraphics[width=0.49\textwidth]{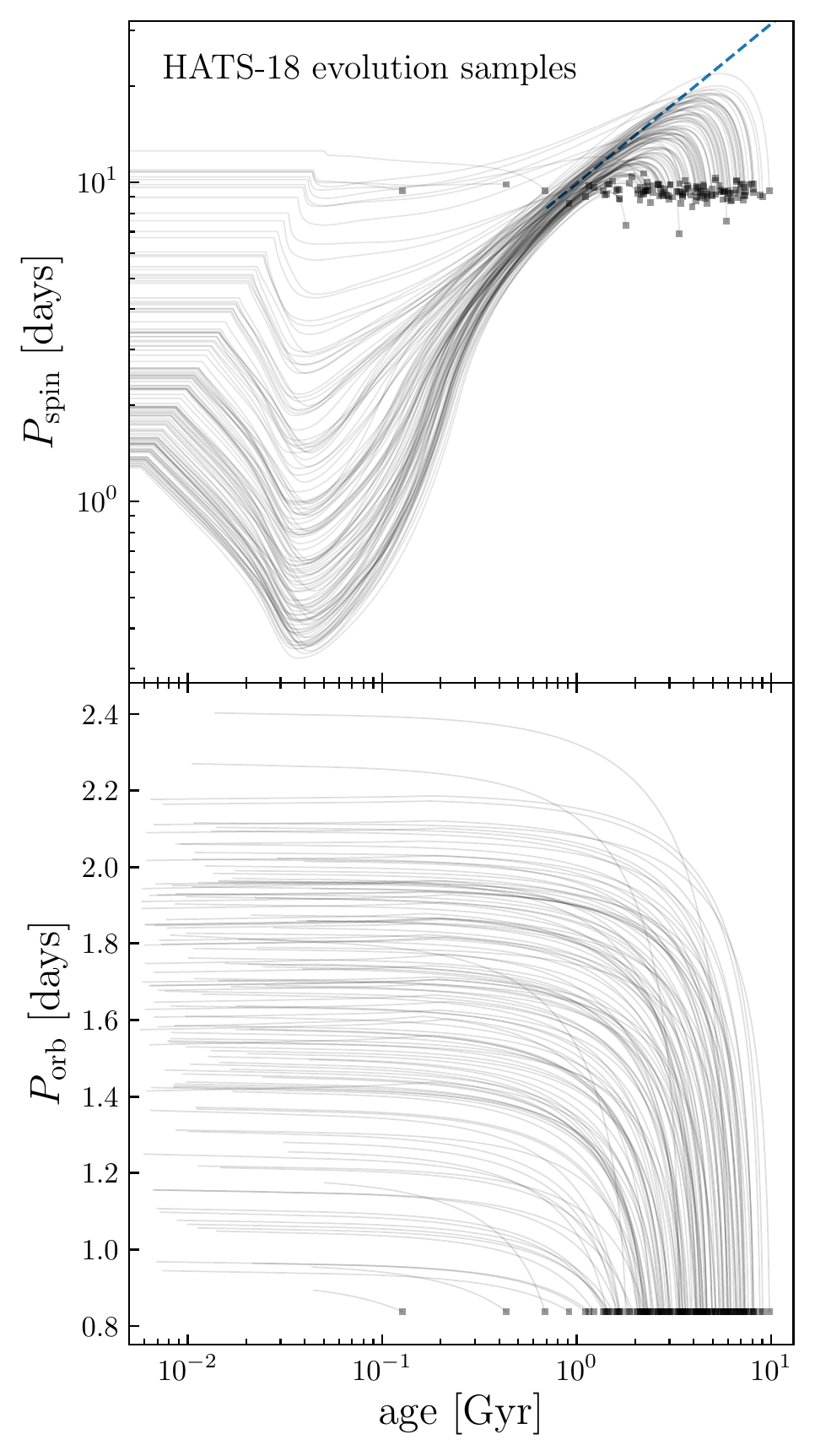}
	\end{center}
	\caption{{\it Top.}---Possible histories of the spin period of
          the HATS-18 system, based on our model for tidal evolution.
          Each curve shows the calculated evolution for a set of
          system parameters drawn randomly from distributions
          representing the observational uncertainties.  The squares
          are random samples of the current spin period and system
          age. The dashed line is the Skumanich relation ($P_{\rm
            spin} \propto t^{1/2}$) which would apply in the absence
          of a planet.  {\it Bottom.}---Same, for the orbital period.
          To determine constraints on $Q_\star'$ we marginalize over
          the uncertainties in all the observed parameters.}
	\label{fig:periods_vs_age}
\end{figure}

\subsection{Orbital and Stellar Spin Evolution}

We model each hot Jupiter system using the Planetary Orbital Evolution due to
Tides (\poet{}) code \citep{Penev_Zhang_Jackson_14}. \poet{} has had numerous
upgrades since its initial release. The version used in this work
self-consistently tracks the orbital decay, the accompanying evolution of the
stellar spin period and obliquity, the loss of angular momentum due to the
stellar wind, and the changes in the star's radius and moment of inertia over
the course of stellar evolution.  The tidal dissipation rate is allowed to be an
arbitrary prescribed function of the system properties.  For the time being,
\poet{} is restricted to circular orbits and does not allow for any thermal
evolution of the planet. For simplicity, we assumed zero obliquity throughout
most of this study.  Some preliminary investigations of obliquity evolution are
described in \S~\ref{sec:discussion}.

\paragraph{Orbital evolution}
The orbital evolution equations are those specified by \citet{Lai_12}. These
equations are general enough to accommodate any prescribed frequency-dependence
of the tidal dissipation rate, and \poet{} also has this flexibility.

\paragraph{Stellar spin evolution}
Our model for stellar spin-down is similar to that used by
\citet{Irwin_et_al_07}, and was described in detail by
\citetalias{Penev_et_al_16}.  In brief, the star is divided into a
radiative core and a convective envelope which can have different rotation
periods. The convective zone loses angular momentum due to a magnetized wind.
Friction acts to gradually torque the two zones toward synchronous rotation.  
We
adopt the coupling timescale for core-envelope angular momentum exchange from the
models of \citet{Gallet_Bouvier_15}, and wind strength parameters from
\citet{Irwin_et_al_07}.  Recently \citet{vanSaders_et_al_16} presented evidence
for a possible loss of spin-down efficiency at long spin periods; while
interesting, this does not
affect the results of our study, since none of the systems that give meaningful
constraints on $Q'_\star$ have stars that spin so slowly.

\paragraph{Stellar evolution}
The orbital and stellar spin evolution depend on the stellar radius and mass,
the moments of inertia of the radiative and convective zones, and on the
location of the core-envelope boundary.  To track the evolution of these
quantities over time, we generate a grid of isochrones using Modules for
Experiments in Stellar
Astrophysics~\citep[MESA;][]{Paxton_et_al_11,Paxton_et_al_13, Paxton_et_al_15}.
We use the solar-tuned inlists from the MESA Isochrones and Stellar Tracks
(MIST) project \citep[][]{Choi_et_al_16, Dotter_MIST_16}. The MIST isochrones
have been validated over a wide range of masses, metallicities, and phases of
stellar evolution, by comparison to existing databases.  We downloaded the
relevant input files and, with minor modifications, computed grids of stellar
evolutionary models spanning all the relevant parameters of our
sample.\footnote{Input files retrieved 2016-09-20 from
    \url{http://waps.cfa.harvard.edu/MIST/data/tarballs_v1.0/MESA_files.tar.gz}.
    As a check on the models, we were able to reproduce the results of
    \citet{Choi_et_al_16}. Our minor modifications were to use the fine-tuned
protosolar abundances discussed in Sec.~4 of \citet{Choi_et_al_16}, rather than
the published grids. We opted for those abundances for greater consistency with
the helioseismic models of the Sun \citep{Christensen-Dalsgaard_et_al_96}.} It
was not possible to simply use the published model outputs, since they did not
include the evolution of the convective and radiative moments of inertia or of
the location of the boundary between the two zones. To compute the two moments
of inertia and the location of the boundary, we identified the tachocline using
the mixing types of each radial cell, and then performed the moment-of-inertia
integral using the density profile $\rho(r)$ at each time step.  We then
interpolated over stellar metallicity and mass to find the stellar properties
corresponding to a given planet-star system at any time in its evolution (see
\citealt{Penev_Zhang_Jackson_14}, Appendix C).

\paragraph{Sub and super-synchronous rotation}
All the stars in our sample have spin periods that are longer than the planet's
orbital period, but in our calculations we do not assume that this was always
the case.  For some systems in our sample, the early evolution goes through a
brief period during which the star spins super-synchronously, and thus the
direction of orbital and stellar spin evolution is reversed. In practice this
does not have a significant effect on the results, because the time intervals of
super-synchronous rotation are brief, and more generally because of the lack of
sensitivity of the current-day parameters to the initial spin period (as
explained below).

\paragraph{Spin-orbit locking}
\poet{} follows all of the relevant physics of sub- and super-synchronous tidal
coupling and the changes in the star's moment of inertia over the course of
stellar evolution, including the possibility that the star-planet system is
temporarily locked in spin-orbit synchrony.  Indeed, in the tidal evolution
calculations to be described below, we observed some cases in which the planets
are able to achieve spin-orbit synchrony with their stars.  But the synchrony is
almost always temporary.  As the star continues to lose angular momentum through
its wind, its tidal bulge begins to lag behind the planet, torquing the planet
into a lower and faster orbit.  To maintain synchrony the star must therefore be
spun up even faster.  This further increases the rate of angular momentum loss
due to the wind.  Thus there ensues a positive feedback spiral.  Ultimately
there comes a time at which the rate of tidal dissipation is insufficient to
maintain synchrony, and the spin-orbit lock is lost.  This same process operates
in binary star systems.  However, due to the much larger secondary mass, the
timescale for breaking spin-orbit synchrony is three orders of magnitude longer,
typically exceeding the age of the universe.  More details were given by
\citet{Penev_Zhang_Jackson_14} who presented an earlier version of the code.

\subsection{Method of Constraining $Q_\star'$}
\label{sec:method_constraining_qstar}

We selected all the known transiting planets from the NASA exoplanet
archive\footnote{\url{exoplanetarchive.ipac.caltech.edu/}} for which
the planet mass exceeds $0.1\,M_{\rm Jup}$, the orbital period is shorter than
3.5\,days, and the star has an effective temperature below 6100\,K.
The effective temperature cut-off ensures that only stars with surface
convective zones are included in the analysis.  We decided to focus on
these ``cool'' stars because both theoretical expectations and
observations suggest the dominant tidal dissipation mechanism, and
hence its efficiency, is dramatically different for stars without a
significant surface convective zone. The restrictions on planet mass
and orbital period were designed
to select systems with strong tides, for which the host stars are most likely
to have been measurably spun up. This resulted in an initial sample of
\Ntotalattempted\ systems.

To model the coupled orbital and spin evolution of each system, we assume that
$Q'_\star$ is constant in time, within a given system.  However, each system is
allowed to have an independent value of $Q'_\star$.  For each system we asked
which value of $Q'_\star$ leads to enough spin-up to explain the observed
rotation rate at the present age, starting from an initial condition compatible
with observations of single stars of the same mass within young clusters. Some
of the key parameters, such as age and spin period, are subject to large
uncertainties. An important part of the work was marginalizing over those
uncertainties.  We use a Monte Carlo technique.  The likelihood is computed as
follows:
\begin{enumerate}
    \item Draw random samples for the radial-velocity semi-amplitude, transit
        characteristics (period, depth, and duration), stellar effective
        temperature, metallicity, and spin period from distributions based on
        the observational uncertainties.  For each parameter we adopt a Gaussian
        probability distribution centered on the value quoted in the literature,
        with a standard deviation given by the quoted 1$\sigma$ uncertainty.
    \item From these samples, compute the posterior distributions for the
        stellar and planetary mass, and the system age (based on the MESA
        models).
    \item Draw a random value for $\log_{10}Q_\star'$ from a uniform
        distribution between 4--9.
    \item Draw a random value for the initial stellar spin period, based on the
        measured spins of stars of similar mass from the Pleiades and M\,50
        clusters (which have an approximate age of 133\,Myrs).
    \item Determine the initial orbital period for which the evolution computed
        by \poet{} brings the system to the currently observed orbital period at
        the system age chosen in step 2.
    \item Compute the likelihood by comparing the observed stellar spin period
        to the calculated spin period at the current age that is predicted by
        the orbital evolution model. Stellar spin periods are taken from
        photometric measurements whenever available, and otherwise from $v\sin
        i$ measurements. The uncertainty distribution for the stellar spin is
        assumed to be Gaussian, with an appropriate mean and standard deviation.
\end{enumerate}

Thus, for each system, the posterior probability distribution is constructed for
$\log_{10}(Q'_\star)$, accounting for all the observational uncertainties.  As
noted above, we assume the orbit to be circular, we neglect obliquity evolution,
and we also neglect the uncertainties in the spin-down model parameters.
Fig.~\ref{fig:periods_vs_age} shows some examples of the calculated time
evolution of the stellar spin and planetary orbital period for the HATS-18
system.  In the top panel, the dashed line shows the Skumanich relation ($P_{\rm
spin} \propto t^{1/2}$) which would apply in the absence of a planet.  The
observed present-day spin period (represented by the squares in
Fig.~\ref{fig:periods_vs_age}) is lower than one would expect without the tidal
torque from a hot Jupiter.
          
The top panel of Fig.~\ref{fig:periods_vs_age} shows that the broad initial
distribution of stellar spin periods collapses to a much narrower range after
$\approx$500\,Myr of evolution.  This generic feature of stellar spin-down is
caused by the angular momentum loss due to the magnetized stellar
wind~\citep{schatzman_theory_1962,weber_angular_1967,skumanich_time_1972}: more
rapidly rotating stars experience stronger magnetic braking.  At least
initially, the presence of a HJ does not affect this well-known feature of
stellar spin evolution.  In other words, the relation between stellar mass, spin
period, and age that is observed in open clusters is preserved for a few hundred
Myrs after the zero-age main sequence.  However, once the host star starts
spinning slowly enough (in Fig.~\ref{fig:periods_vs_age}, at $P_{\rm
spin}\approx 15\,{\rm days}$, ${\rm age}\approx 1-2\,{\rm Gyr}$), the rate of
angular momentum loss through the magnetized wind is matched by the rate of
angular momentum gain from tidal interactions with the planet.  At this time,
the star begins spinning up, and it continues to gain angular momentum until it
reaches its present-day spin period.

The stellar spin at an age of $\approx 500\,{\rm Myr}$ hardly depends at all on
the spin at earlier times (Fig.~\ref{fig:periods_vs_age}).  This implies that
the choice of which young clusters to use as references for the initial rotation
period is relatively unimportant.  Because the early history of the stellar spin
is quickly forgotten, the inferred value of $Q'_\star$ can be loosely
interpreted as a measure of the present-day value of $Q'_\star$, i.e., even if
in reality $Q'_\star$ has changed over time, in contradiction with the premise
of our evolution calculations, we will infer a value of $Q'_\star$
representative of the present-day value (as shown quantitatively in
\S~\ref{subsec:empirical_validation}).

For the same reason, our results for $Q'_\star$ do not depend on the unknown
events that lead to the formation of the HJs, provided they take place within
the first few hundred million years of the system.  Whether hot Jupiters arrive
in their tight orbits within the first million years through disk migration, or
200~Myr later through some other process such as high-eccentricity migration,
the final stellar spin is hardly affected.  We note, though, that the proposed
high-eccentricity mechanisms for hot Jupiter formation \citep[see,
e.g.][]{Fabrycky_Tremaine_07} operate on a very wide range of timescales, up to
billions of years. If the long-timescale mechanisms were dominant then the
actual time for tidal evolution would be shorter than is assumed in our
calculations.  This caveat should be kept in mind, although many studies have
concluded that the high-eccentricity mechanism is unlikely to be the dominant
mechanism for producing HJs \citep[see,
e.g.][]{Naoz+2012,CridaBatygin2014,Dawson+2015,Petrovich_2015, Ngo+2015,
Ngo+2016, SchlaufmanWinn2016}.

\section{Results}
\label{sec:results}

\begin{figure}[!t]
	\begin{center}
		\includegraphics[width=0.49\textwidth]{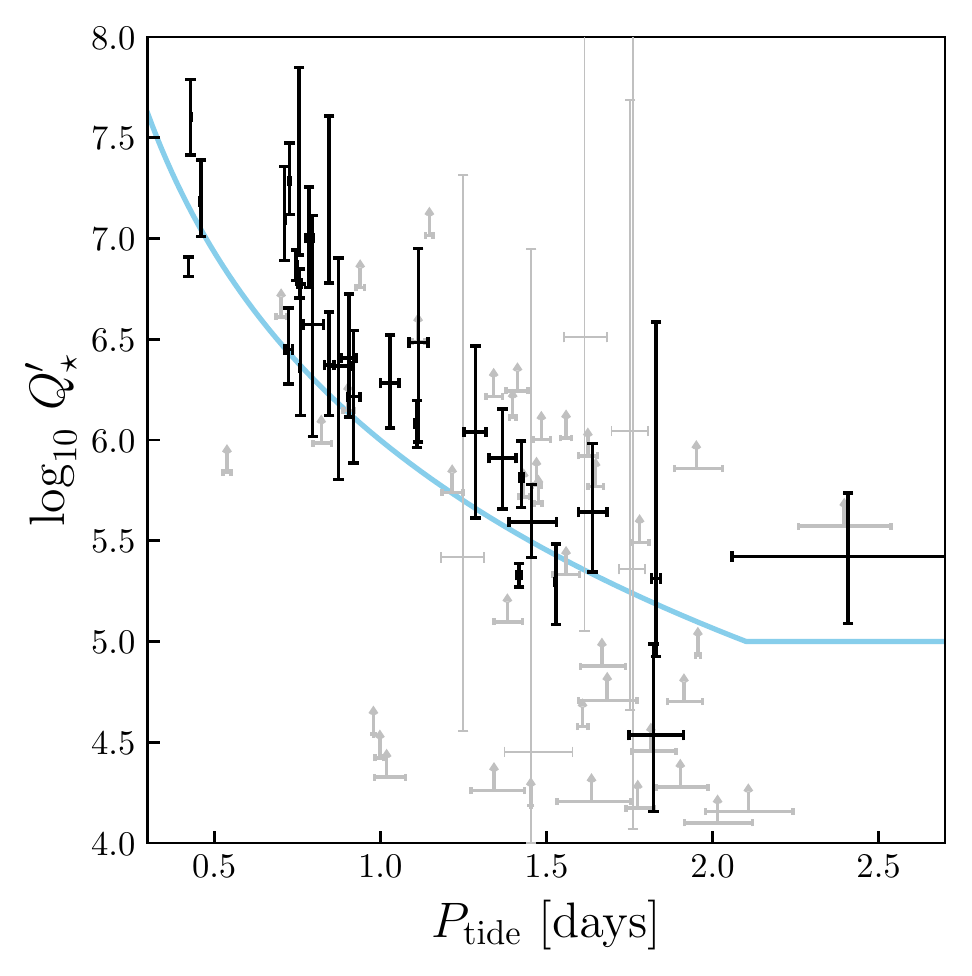}
		
	\end{center}
	\caption{
	Evidence for the frequency-dependence of $Q'_\star$.
	Shown are the values of $Q'_\star$ inferred for each system,
	as a function of the tidal forcing period.
	Black points are cases for which $Q_\star'$ was bounded to within two orders of magnitude.
	Thinner gray symbols are cases for which the two-sided limits span more than two orders of magnitude.
	Gray arrows indicate lower limits.
	The blue curve is a saturated power-law fit to the points with two-sided limits
	(Eq.~\ref{eq:Qstar_fit}).
	}
	\label{fig:logQ_vs_Ptide}
\end{figure}
\begin{figure}[!t]
	\begin{center}
		\includegraphics[width=0.49\textwidth]{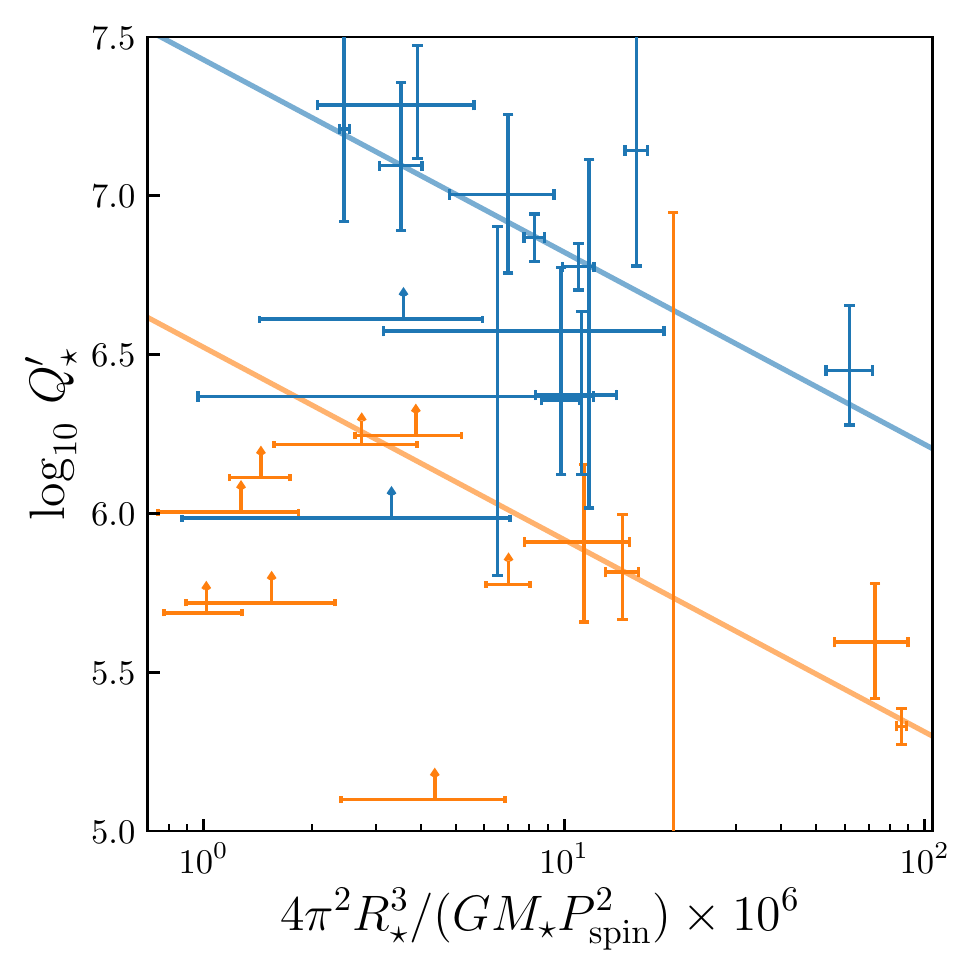}
	\end{center}
	\caption{
		Tentative evidence for a secondary dependence of $Q'_\star$ on the stellar spin rate.
		Shown are the values of $Q'_\star$ inferred for selected systems,
  	as a function of a dimensionless spin parameter. The blue symbols 
		correspond to systems with tidal periods from 0.7 to 0.9\,days, 
		and the orange symbols are for tidal periods from 1.3 to 1.5\,days.
		The blue curve (a power-law dependence) is a fit to blue 
		symbols with two-sided limits.
		For the orange line, only the offset was fitted, while the slope was forced to be 
		the same as the blue line.
	}
	\label{fig:logQ_vs_spin_parameter}
\end{figure}

Of the \Ntotalattempted\ systems in the sample, our procedure led to two-sided
limits on $Q_\star'$ for \Ntwosidedtotal\ systems.  In another \Nlowerlim\
cases, it was possible to derive a lower bound on $Q_\star'$. In the remaining
cases the data did not lead to meaningful constraints.  Table~\ref{tbl:results}
summarizes the quantitative results for each system.
Fig.~\ref{fig:logQ_vs_Ptide} shows the inferred value of $\log_{10}(Q'_\star)$
for each system, as a function of the tidal forcing period.  The tidal forcing
period is one-half of the orbital period of the planet in a reference frame
rotating with the stellar spin:
\begin{equation}
    \label{eq:ptide}
    P_{\rm tide} \equiv \frac{1}{2(P_{\rm orb}^{-1} - P_{\rm spin}^{-1})}.
\end{equation}

We see a clear trend toward larger $Q_\star'$ with decreasing $P_{\rm tide}$.
The blue line is a simple function that fits the period dependence,
\begin{equation}
    Q_\star'(P_{\rm tide})
    =
    \max\left[\frac{10^{6.0}}{(P_{\rm tide}/{\rm days})^{3.1}},10^{5}\right].
\label{eq:Qstar_fit}
\end{equation}
The minimum value of $Q_\star'=10^5$ was imposed because it agrees with the
constraints obtained by \citet{Milliman_et_al_14} (M14, hereafter) based on
observations of the eccentricity-period relation for binary stars in open
clusters. This minimum value is also consistent with the results from the
longest-period systems in our sample.  

About half of the systems for which two-sided constraints were
obtained have tidal periods in the relatively narrow range $P_{\rm
  tide}=$~0.7--0.9~day.  This allows us to check for any secondary
dependence of $Q_\star'$ on other system parameters, at fixed period.
We looked for trends with the planet mass, stellar mass, and stellar
spin.  Of these, the spin showed the strongest correlation with
$Q_\star'$, with a formal false alarm probability of
$5.5\times10^{-4}$.  Fig.~\ref{fig:logQ_vs_spin_parameter} displays
this result as a correlation between $Q_\star'$ and the dimensionless
parameter
\begin{equation}
  \left( \frac{\Omega_{\rm spin}}{\Omega_{\rm crit}} \right)^2 = \left[ \frac{(2\pi/P_{\rm spin})} {\sqrt{GM_\star/R_\star^3} } \right]^2 =
  \frac{4\pi^2 R_\star^3}{GM_\star P_{\rm spin}^2},
\end{equation}
where $\Omega_{\rm crit}$ is the critical angular velocity for breakup
due to centrifugal forces.  In this figure the blue points are the
systems with $P_{\rm tide}=$~0.7--0.9\,d. The orange points are a
separate set of systems with $P_{\rm tide}$ in the range from
1.3-1.5~days.  By itself, the orange collection of points would not be
sufficient to establish that a correlation exists, but they are at
least consistent with the same slope that fits the blue points.

\subsection{Self-consistency of results}
\label{subsec:empirical_validation}

\begin{figure}[t!]
    \begin{center}
        \includegraphics[width=0.49\textwidth]{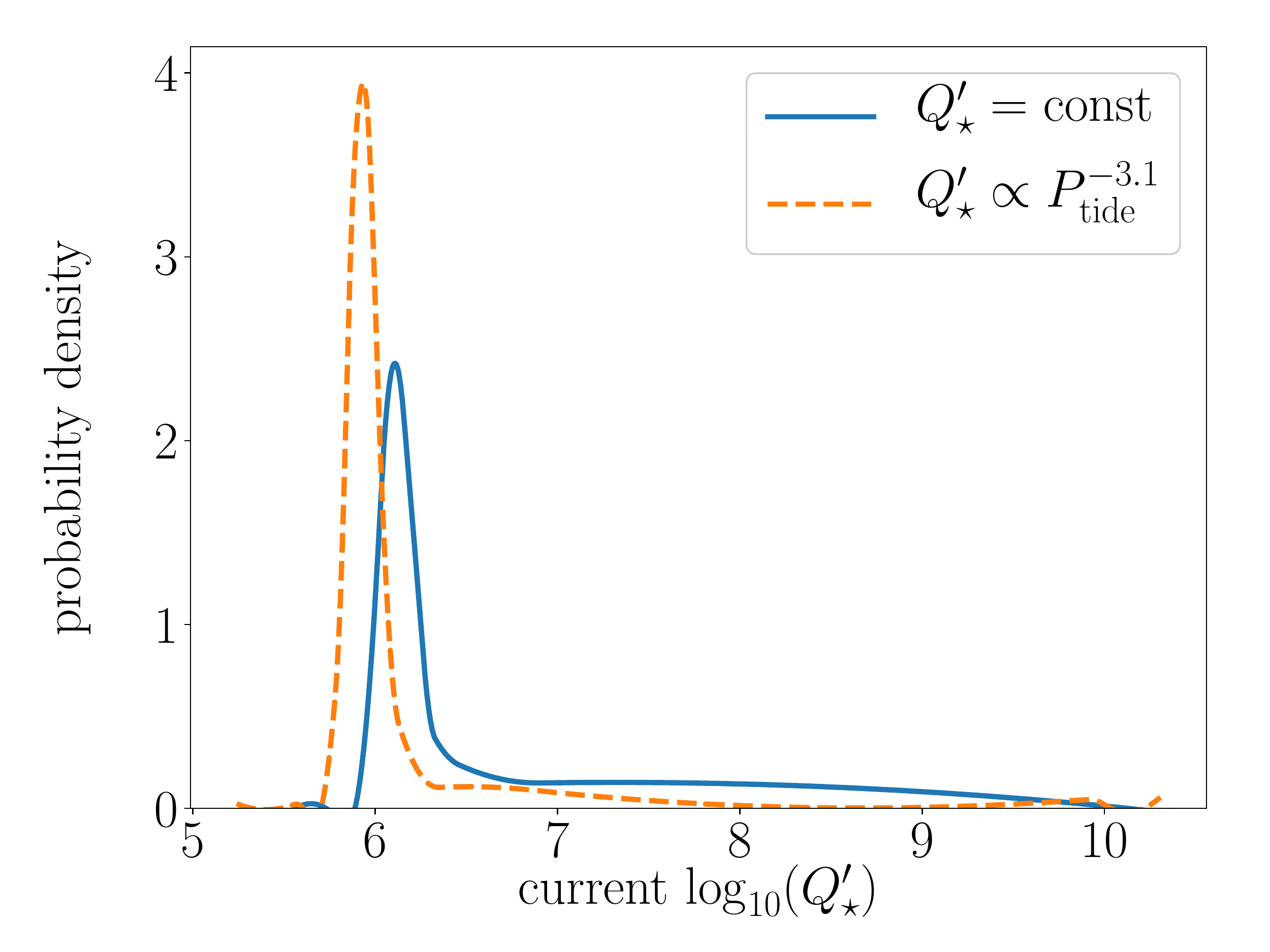}
    \end{center}
    \caption{
        Probability distributions for the present day $Q'_{\star}$ for WASP-50,
        computed under the assumption of a frequency-independent $Q'_{\star}$
        (solid), and the frequency-dependent prescription for $Q'_{\star}$ found
        in \refsec{results} (dashed). The constraints are similar, supporting
        the claim that our results are reliable even though they were derived
        under the assumption that $Q'_{\star} =$~constant for each system.
    }
    \label{fig:robustness}
\end{figure}

\begin{figure}[t!]
    \begin{center}
        \includegraphics[width=0.49\textwidth]{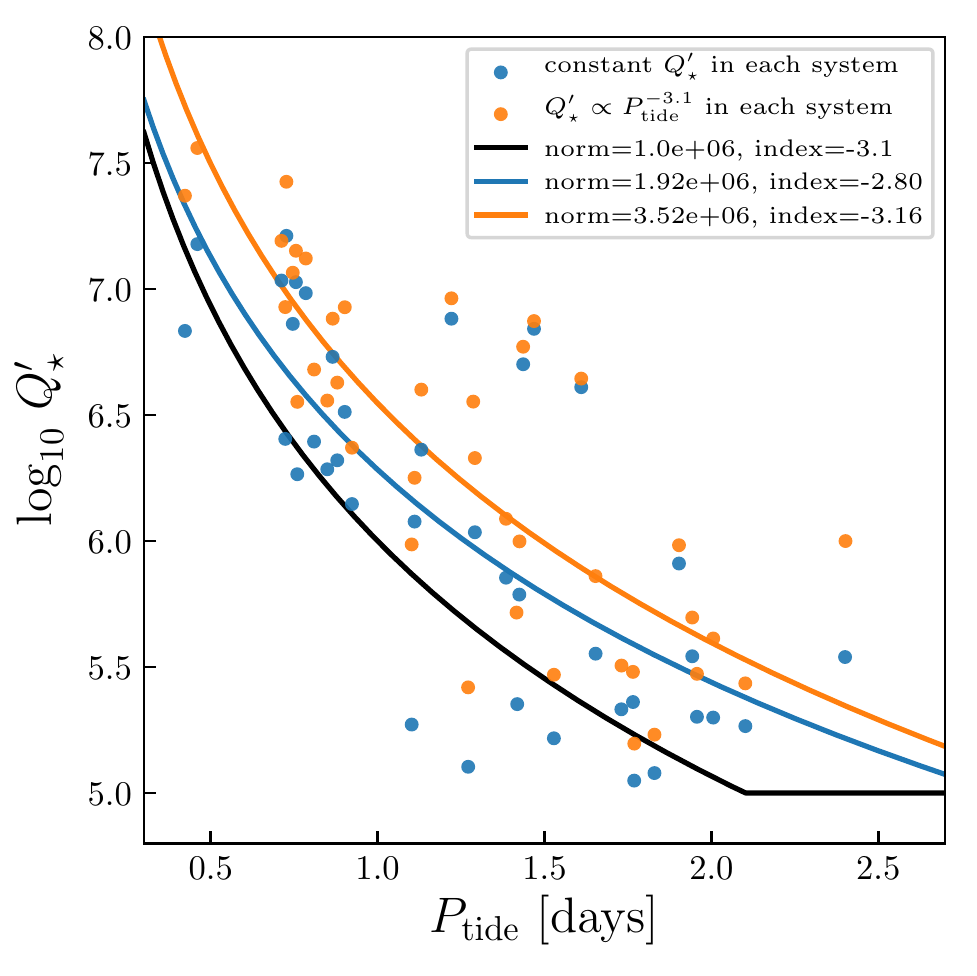}
    \end{center}
    \caption{
        Estimate of how the assumed $Q_\star'(P_{\rm tide})$ dependence affects
        the derived $Q_\star'(P_{\rm tide})$ relation.  The displayed points
        come from the method described in
        Sec.~\ref{subsec:empirical_validation}.  The blue points assume
        $Q_\star'(P_{\rm tide})={\rm constant}$, and the fit (blue line) weights
        these points uniformly.  The orange points assume $Q_\star'(P_{\rm
        tide})=Q_0 P_{\rm tide}^{-3.1}$, and solves for an appropriate $Q_0$
        (see text).  The orange line weights orange points equally.  The black
        line (Eq.~\ref{eq:Qstar_fit}) is the same as from
        Fig.~\ref{fig:logQ_vs_Ptide}, which was found by applying the method
        described in Sec.~\ref{sec:method_constraining_qstar}.  Regardless of
        the assumed frequency scaling for each system, and regardless of the
        assigned weights each system receives, stars that are forced faster
        dissipate less efficiently.
    }
    \label{fig:robustness_2}
\end{figure}

Our calculations assumed that $Q_\star'$ is a constant in time, while the
results suggest that it actually depends sharply on frequency.  To investigate
the systematic error associated with this inconsistency, we repeated the
calculations but this time requiring $Q_\star' \propto P_{\rm tide}^{-3.1}$,
with a normalization that is specific to each system.  However, since it would
require additional months of computing time to perform another iteration of the
entire calculation, we confirmed that the systematic errors are relatively small
through two more tractable calculations.

First we repeated the analysis of the WASP-50 system, which gives the tightest
constraint on $Q'_\star$ near the middle of the period range of our sample.  We
found that the inferred value of $Q'_\star$ differs by 0.2~dex from the case in
which we assumed $Q'_\star$ to be a constant, as shown in
Fig.~\ref{fig:robustness}.  Similar tests, with similar results, had already
been performed by \citet{Penev_et_al_16} for the two shortest-period systems in
the sample, HATS-18 and WASP-19.

Next we repeated the analysis for all the most important members of the sample,
but without the computationally expensive marginalization over all of the
observational uncertainties.  We analyzed all the systems that gave two-sided
limits in Sec.~\ref{sec:method_constraining_qstar}.  For each system we took the
initial spin period to be the median of the measured periods in the
Pleaides/M\,50 samples, and adopted stellar and planetary parameters based on
the best-fit values reported in the literature.  Then, under the assumption
\begin{equation}
    Q_\star' = 
    {\rm max}\left[
        Q_0\times \left(\frac{P_{\rm tide}}{{\rm days}}\right)^{-3.1},\ 10^5
    \right],
    \label{eq:variable_Qstar}
\end{equation}
we determined the values $Q_0$ and the initial orbital period that lead to a
good match between the calculated and observed values of the present-day orbital
and spin periods (or $v \sin i$, when the period is not available).

The orange points in Fig.~\ref{fig:robustness_2} show the results of this
exercise.  An unweighted power-law fit to the collection of results gives
$Q_\star' \propto P_{\rm tide}^{-3.1}$.  The blue points show the results of
performing exactly the same procedure under the assumption $Q_\star'=$~constant
for each system; here the power-law fit gives $Q_\star' \propto P_{\rm
tide}^{-2.8}$.  The overall normalizations differ by about a factor of two.  We
consider this to be good agreement, and a sign that no further steps toward
total self-consistency are warranted.  The results also show that the results
reported in Fig.~\ref{fig:logQ_vs_Ptide} and Table~\ref{tbl:results} are subject
to systematic uncertainties of about a factor of two, in addition to the
reported statistical uncertainties.

We had anticipated that the inferred present-day $Q_\star'$ values would be
insensitive to the system's history because of the strong dependence of magnetic
braking on the stellar spin frequency.  The faster the spin, the faster angular
momentum is lost.  In the absence of tides, this relationship is what causes the
spin rates of initially fast and slow rotators to converge over time, leading to
the well known ``gyrochronological'' relationship between age and rotation rate.
In the case of tidal spin-up, spinning up the star at early times has hardly any
effect on the present-day properties because the excess angular momentum is
quickly lost due to the enhanced magnetic braking.  As a result, any excess spin
that can be detected now is a measure of only the most recent history of angular
momentum deposition.

\section{Discussion}
\label{sec:discussion}

\begin{figure*}[t!]
	\plotone{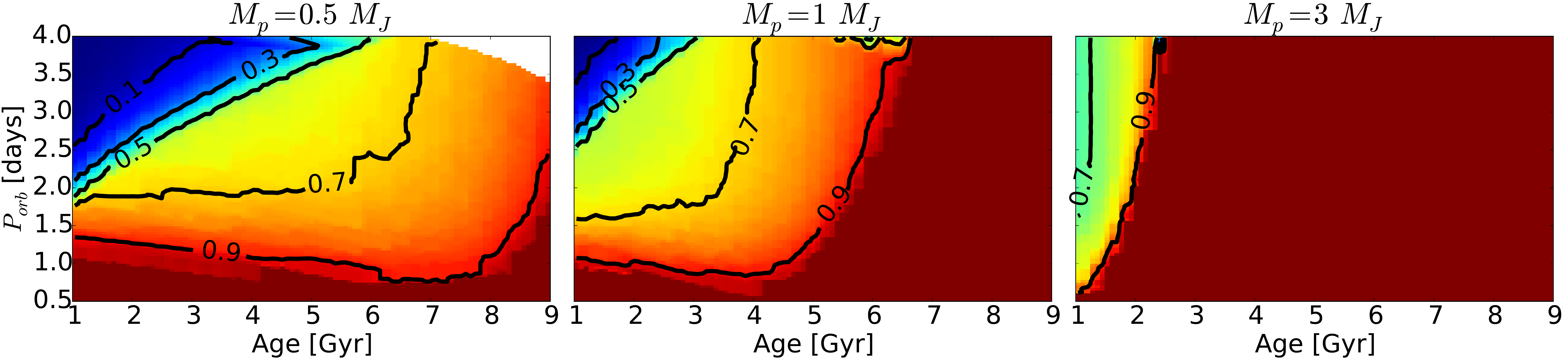}
	\caption{
		Fraction of initial spin--orbit orientations that end up aligned by a
		given age and with a given orbital period due to orbital evolution,
		assuming the frequency-dependent $Q'_\star$ given by Eqn.~\ref{eq:Qstar_fit}.
		The host star is $1~M_\odot$ in all cases. The spin-down
		parameters are taken from \citep{Irwin_et_al_07}, and the three panels
		are for planet masses of 0.5, 1 and 3~$M_{\rm Jup}$.
	}
	\label{fig:tidal_alignment}
\end{figure*}

\subsection{Other Empirical Constraints on $Q'_\star$}

\citet{Meibom_Mathieu_05} (hereafter, MM05) and
\citet{Milliman_et_al_14} (M14) examined the eccentricity distribution
of binaries within open clusters of differing ages.  For each cluster
they tried to find the orbital period $P_{\rm circ}$ below which
binaries have been tidally circularized and above which they have not
yet circularized. They determined $Q'_\star$ by fitting the trend of
$P_{\rm circ}$ with age, thereby assuming $Q'_\star$ to be the same
for all stars, ages, and tidal periods. These authors argued that
significant circularization occurs on the main-sequence, requiring
$Q_\star' \lesssim 10^{5.5}$. However there are some puzzling results
that call into question the premises of this method. The observed
$P_{\rm circ}$ is not a monotonic function of age: Praesepe and the
Hyades (630~Myr) have smaller values of $P_{\rm circ}$ than the
younger clusters. Also notable is that NGC~6819 (2.5 Gyr) is a large
outlier from the overall main-sequence circularization trend. Thus,
three out of the seven clusters used in that study are inconsistent
with the claimed result. In any case, the existence of a particular
$P_{\rm circ}$ must be an oversimplification because all the clusters
show a significant overlap in orbital period between circularized and
highly eccentric systems \citep[cf.][]{Mazeh_08}. Furthermore, taking
the conclusions of MM05 and M14 at face value, one would predict
unrealistically rapid rates for the orbital decay of HJs. To avoid
destroying too many HJs, \citet{Penev_et_al_12} (P12) argued that
$Q_\star'$ must exceed $10^7$.

Tidal dissipation also appears to have shaped the eccentricity
distribution of short-period giant planets.  By integrating the tidal
evolution equations backward in time for the shortest-period hot
Jupiters, and requiring their initial eccentricity distribution to
match that of warm Jupiters (longer-period planets of similar mass),
\citet{Jackson_Greenberg_Barnes_08a} found $Q_p'\sim10^{5.5}$ for the
planets, and $Q_\star'\sim10^{6.5}$ (with a large uncertainty) for the
stars.  Similar results were obtained by \citet{Husnoo_et_al_12}.  On
the other hand, \citet{Hansen_10} interpreted the same data using a
different formalism, in which the tidal dissipation rate was allowed
to vary as expected with the size of the body and its orbital
distance. He found lower dissipation rates of $10^7<Q_p'<10^8$ and
$10^7<Q_\star'<10^9$.  Most recently \citet{Bonomo_et_al_17}, using
the most up-to-date observations of transiting HJs and simple
timescale arguments, found that $Q_\star'$ has to exceed $10^6$ or $10^7$ (depending on the
system) and $10^5 < Q_p' < 10^9$.

Our results suggest a possible reconciliation of all of these
confusing and seemingly inconsistent results.  At the shortest-period
end, we find $Q_\star' \sim 10^7$ which is consistent with the
previous study of tidal inspiral by \citet{Penev_et_al_12}.  For the
longest periods in our sample we found $Q_\star' \sim 10^5$, which is
compatible with the trend of circularization period versus age in
cluster binaries seen by \citet{Meibom_Mathieu_05} and
\cite{Milliman_et_al_14}, and with most of the constraints based on
hot and warm Jupiters \citep{Jackson_Greenberg_Barnes_08a,
  Husnoo_et_al_12}.  This leaves only the higher $Q_\star'$ values
found by \citet{Hansen_10}.  Regarding those, we note that the results
hinge on the observed non-circular orbits of WASP-14b, XO-3b and
HAT-P-2b.  Those particular planet-hosting stars are right on the
boundary between being convective or radiative near the surface.  They
may have radiative envelopes, or at least exist in a different, much
less efficient, regime of tidal dissipation due to the extremely thin
convective zone.

\subsection{Tidal Alignment of Hot Jupiter Systems}
\label{sec:tidal_alignment}

Observations have revealed a wide range of obliquities for the host
stars of HJs, including very well-aligned systems
\citep[e.g.][]{Winn_et_al_06}, strongly misaligned systems
\citep{Hebrard_et_al_08}, and even polar and retrograde orbits
\citep{Winn_et_al_09, Narita_et_al_09, Queloz+2010}. The
basis for most of these results is the Rossiter-McLaughlin
effect, a spectroscopic distortion that occurs during transits. A
pattern has emerged from these results: the host stars cool enough to
have outer convective zones tend to have low obliquities, while hotter
stars show a broader range of obliquities \citep{Winn_et_al_10,
  Albrecht_et_al_12}.

One possible interpretation is that HJs generally form with a broad
range of orbital orientations, and that only the cool stars are able
to tidally re-align the system due to the enhanced tidal dissipation
rate associated with outer convective zones. A problem with this
interpretation is that simple tidal models predict that stellar
realignment should occur on the same timescale as orbital decay,
making it unlikely that we would ever observe a re-aligned hot Jupiter
system. Whether or not there are viable mechanisms for re-alignment
that are faster than orbital decay is an outstanding theoretical
question \citep{Lai_12,RogersLin13,LiWinn16}.

Our finding of a strongly frequency-dependent $Q_\star'$ is relevant to this
issue.
As elucidated by \citet{Lai_12}, the timescales for obliquity damping and 
orbital decay can be different, provided that a system is misaligned.
In particular, the obliquity can be affected by tidal waves with much lower 
frequency ($2\Omega_{\rm spin}$) than the ones causing tidal decay ($\sim 2 
\Omega_{\rm orb}$).
The frequency dependence of $Q_\star'$ suggested by our study may allow a way for the stellar obliquity to be 
damped to a low value, before the orbit decays.
This increases the chance of discovering a low obliquity system before the 
planet is destroyed.
Fig.~\ref{fig:tidal_alignment} shows the results of our preliminary
investigation of this issue, using nominal system parameters rather than the
actual parameters of the known systems.  We considered a solar-mass star, a
planet with one of three possible masses, and an initial obliquity chosen
from an isotropic distribution.  The contours in this figure show the
fraction of planet--star systems for which the obliquity damps to below
5$^\circ$ before the planet is tidally destroyed, as a function of the age
and observed orbital period.  The evolution was calculated using the
prescription for $Q'_\star$ given by Eq.~\ref{eq:Qstar_fit}.

The results --- obtained without tuning any free parameters --- show
that obliquity damping is achieved over much of the relevant parameter
space. This is an enticing result: it suggests that the
frequency-dependence of tidal dissipation which we have inferred from
stellar spin-up could be sufficient to explain the high degree of
alignment observed for planets around stars with surface convective
zones without driving the planet into the star through orbital decay.
We intend to investigate this issue more rigorously in the future,
with a detailed system-by-system evaluation and a comparison with the
measured obliquity distribution.  If successful, then we might be able
to explain the observed obliquity pattern while also providing
information about the initial distribution of stellar obliquities.

\subsection{Conclusions \& Caveats}

By modeling the tidal evolution of a large sample of hot Jupiters orbiting
main-sequence stars with external convection zones, we have derived limits on
the dissipation parameter $Q_\star'$ and found evidence for a strong frequency
dependence of this parameter (Fig.~\ref{fig:logQ_vs_Ptide}), while systems with
similar tidal forcing periods led to similar results for $Q'_\star$.  There is
also a possible correlation between $Q'_\star$ and the stellar spin frequency,
at constant tidal period (Fig.~\ref{fig:logQ_vs_spin_parameter}).  At the
shortest tidal periods, $Q'_\star$ matches the \citet{Penev_et_al_12}
constraints from exoplanet inspiral, while at the longest periods $Q'_\star$
matches the \citet{Milliman_et_al_14} constraints from open cluster binary star
circularization, thus reconciling those seemingly inconsistent results.  We also
showed that with no further free parameters, our prescription for the
frequency-dependent $Q'_\star$ might naturally explain the fact that HJs around
cool stars tend to be in nearly perfectly equatorial orbits around their stars,
while those around hotter stars have a broader range of orbital orientations.

We can think of many ways to improve on our analysis, mainly for
self-consistency.  It might be warranted to attempt a hierarchical Bayesian
analysis, in which the constant-$Q'_\star$ model is replaced with
\begin{equation}
    Q_{\star}' = Q'_0 
    \left( \frac{P_{\rm tide}}{\mathrm{day}} \right)^\alpha
    \left( \frac{\Omega_{\rm spin}}{\Omega_{\rm crit}} \right)^\beta,
\end{equation}
where $P_{\rm tide}$ is defined by Eq.~\ref{eq:ptide}, and the parameters
$Q'_0$, $\alpha$ and $\beta$ are determined by modeling all of the systems
simultaneously.  As in Section~\ref{sec:results}, the purpose of the second
factor is to allow for an explicit dependence on stellar spin rate, even for a
fixed tidal forcing period.  Such a dependence could arise from rotationally
induced changes in the structure of the star and the oscillations it supports.
Our initial explorations (Sec.~\ref{subsec:empirical_validation}) suggest that
moving toward this level of self-consistency could affect the inferred
normalization $Q'_0$ by about a factor of two, and the exponent by about 0.2.
We also found tentative evidence for a nonzero value of $\beta$.  In principle,
if $\beta$ turns out to be too large, the spin-dependence of the tidal
dissipation rate could overwhelm the spin-dependence of magnetic braking.  This
would lead to a more sensitive (and problematic) dependence of our results on
the initial conditions.  The current data, though, give no suggestion that
$\beta$ is large enough for this situation to occur.

Another consideration is that if the tidal dissipation in cool stars is
responsible for the spin-orbit alignment of HJ systems, then self-consistency
requires a consideration of the initial obliquity distribution of the stars.
Based on the numerical experiments of \citetalias{Penev_et_al_16} (see their
Fig.~10), allowing for a wide range of initial obliquities has the effect of
extending the allowed range of $Q'_\star$ to lower values.

Another limitation of our study is that we neglected the uncertainties
in the parameters of the spin-down model.  For instance, based on
Fig.~8 of \citet{Gallet_Bouvier_15}, we assumed a constant
core-envelope coupling timescale of 10\,Myr.  While these authors
found that to be appropriate for stars with masses comparable to the
Sun, they also found evidence that this parameter should really be a
function of the amount of differential rotation, with faster coupling
for strong differential rotation.  We adopted the value corresponding
to large differential rotation because that is the regime where this
parameter has the strongest effect. However, a more complete treatment
would marginalize over the possible values for this timescale, as well
as the parameters in the model of magnetic braking.  Again we do not expect
the results would change dramatically, since for the two systems
considered by \citetalias{Penev_et_al_16} the results were robust even to
implausibly large changes in the coupling timescale and spin-down
parameters.

Another detail is that the stellar-evolutionary models we relied upon
do not include any effects of stellar rotation. Although stellar
rotation has only a minor impact on the bulk stellar properties for
low-mass stars, this may be worth including for self-consistency in
future studies.

\acknowledgements
JH acknowledges support from NASA grant NNX17AB61G. This research has made
use of the NASA Exoplanet Archive, which is operated by the California
Institute of Technology, under contract with the National Aeronautics and
Space Administration under the Exoplanet Exploration Program.

\appendix
\startlongtable
\begin{deluxetable}{lcccccccc}

\tablecaption{Derived stellar quality factors for all systems with single or 
two-sided constraints.
\label{tbl:results}    }
    
\tabletypesize{\scriptsize}\tablehead{
	\colhead{System} &
	\colhead{Orbital Period} &
	\colhead{Spin Period} &
	\colhead{Stellar Mass} &
	\colhead{Stellar Radius} &
	\colhead{Planetary Mass} &
	\colhead{Planetary Radius} &
	\colhead{$\log_{10}(\mathrm{Q}_\star')$}
\\
	\colhead{} &
	\colhead{[days]} &
	\colhead{[days]} &
	\colhead{[M$_\odot$]} &
	\colhead{[R$_\odot$]} &
	\colhead{[M$_\mathrm{J}$]} &
	\colhead{[R$_\mathrm{J}$]} &
	\colhead{}
}
\startdata
	CoRoT-12 &
	\num{2.83}$\pm$\num{1.3e-05} &
	\num{6e+01}$\pm$\num{6e+01} &
	$\num{1.08}^{+\num{0.08}}_{-\num{0.07}}$ &
	$\num{1.12}^{+\num{0.1}}_{-\num{0.09}}$ &
	$\num{0.917}^{+\num{0.07}}_{-\num{0.065}}$ &
	\num{1.44}$\pm$\num{0.13} &
	$>$\num{5.33}
\\
	CoRoT-18 &
	\num{1.9}$\pm$\num{2.8e-06} &
	\num{5.4}$\pm$\num{0.5} &
	\num{0.95}$\pm$\num{0.15} &
	\num{1}$\pm$\num{0.13} &
	\num{3.5}$\pm$\num{0.38} &
	\num{1.3}$\pm$\num{0.18} &
	$\num{5.59}^{+\num{0.185}}_{-\num{0.177}}$
\\
	CoRoT-2 &
	\num{1.74}$\pm$\num{1e-06} &
	\num{4.5}$\pm$\num{0.5} &
	\num{0.96}$\pm$\num{0.08} &
	\num{0.91}$\pm$\num{0.03} &
	\num{3.47}$\pm$\num{0.22} &
	$\num{1.47}^{+\num{0.042}}_{-\num{0.044}}$ &
	$\num{5.33}^{+\num{0.0565}}_{-\num{0.0582}}$
\\
	CoRoT-29 &
	\num{2.85}$\pm$\num{6e-06} &
	\num{13}$\pm$\num{2.5} &
	\num{0.97}$\pm$\num{0.14} &
	\num{0.9}$\pm$\num{0.12} &
	\num{0.85}$\pm$\num{0.2} &
	\num{0.9}$\pm$\num{0.16} &
	$\num{4.54}^{+\num{0.452}}_{-\num{0.379}}$
\\
	HAT-P-13 &
	\num{2.92}$\pm$\num{1.5e-05} &
	\num{48}$\pm$\num{11} &
	$\num{1.22}^{+\num{0.05}}_{-\num{0.1}}$ &
	\num{1.56}$\pm$\num{0.08} &
	\num{0.851}$\pm$\num{0.038} &
	\num{1.28}$\pm$\num{0.079} &
	$>$\num{6.01}
\\
	HAT-P-20 &
	\num{2.88}$\pm$\num{4e-06} &
	\num{7.28}$\pm$\num{0.5} &
	\num{0.76}$\pm$\num{0.03} &
	\num{0.69}$\pm$\num{0.02} &
	\num{7.25}$\pm$\num{0.187} &
	\num{0.867}$\pm$\num{0.033} &
	$>$\num{5.57}
\\
	HAT-P-23 &
	\num{1.21}$\pm$\num{1.7e-07} &
	\num{6.81}$\pm$\num{0.46} &
	\num{1.1}$\pm$\num{0.05} &
	\num{1.09}$\pm$\num{0.03} &
	\num{2.07}$\pm$\num{0.12} &
	\num{1.37}$\pm$\num{0.09} &
	$\num{6.45}^{+\num{0.205}}_{-\num{0.171}}$
\\
	HAT-P-27 &
	$\num{3.04}^{+\num{5e-06}}_{-\num{6e-06}}$ &
	\num{1e+02}$\pm$\num{1e+02} &
	\num{0.94}$\pm$\num{0.04} &
	$\num{0.9}^{+\num{0.05}}_{-\num{0.04}}$ &
	\num{0.62}$\pm$\num{0.03} &
	$\num{1.04}^{+\num{0.077}}_{-\num{0.058}}$ &
	$>$\num{5.77}
\\
	HAT-P-28 &
	\num{3.26}$\pm$\num{7e-06} &
	\num{3e+02}$\pm$\num{5e+02} &
	\num{1.02}$\pm$\num{0.05} &
	$\num{1.1}^{+\num{0.09}}_{-\num{0.07}}$ &
	\num{0.626}$\pm$\num{0.037} &
	$\num{1.19}^{+\num{0.102}}_{-\num{0.075}}$ &
	$>$\num{5.49}
\\
	HAT-P-36 &
	\num{1.33}$\pm$\num{3e-06} &
	\num{15.3}$\pm$\num{0.5} &
	\num{1.02}$\pm$\num{0.05} &
	\num{1.1}$\pm$\num{0.06} &
	\num{1.83}$\pm$\num{0.099} &
	\num{1.26}$\pm$\num{0.071} &
	$\num{7.29}^{+\num{0.188}}_{-\num{0.167}}$
\\
	HAT-P-37 &
	\num{2.8}$\pm$\num{7e-06} &
	\num{15}$\pm$\num{2.6} &
	\num{0.93}$\pm$\num{0.04} &
	\num{0.88}$\pm$\num{0.06} &
	\num{1.17}$\pm$\num{0.103} &
	\num{1.18}$\pm$\num{0.077} &
	$>$\num{4.71}
\\
	HAT-P-43 &
	\num{3.33}$\pm$\num{1.5e-05} &
	\num{23}$\pm$\num{4.9} &
	$\num{1.05}^{+\num{0.03}}_{-\num{0.04}}$ &
	$\num{1.1}^{+\num{0.04}}_{-\num{0.02}}$ &
	\num{0.662}$\pm$\num{0.06} &
	$\num{1.28}^{+\num{0.057}}_{-\num{0.034}}$ &
	$>$\num{4.70}
\\
	HAT-P-52 &
	\num{2.75}$\pm$\num{9.4e-06} &
	\num{75}$\pm$\num{63} &
	\num{0.89}$\pm$\num{0.03} &
	\num{0.89}$\pm$\num{0.05} &
	\num{0.818}$\pm$\num{0.029} &
	\num{1.01}$\pm$\num{0.072} &
	$>$\num{6.00}
\\
	HAT-P-53 &
	\num{1.96}$\pm$\num{3.9e-06} &
	\num{15}$\pm$\num{2} &
	\num{1.09}$\pm$\num{0.04} &
	$\num{1.21}^{+\num{0.08}}_{-\num{0.06}}$ &
	\num{1.48}$\pm$\num{0.056} &
	\num{1.32}$\pm$\num{0.091} &
	$\num{6.48}^{+\num{0.468}}_{-\num{0.493}}$
\\
	HAT-P-65 &
	\num{2.61}$\pm$\num{3.1e-06} &
	\num{13.3}$\pm$\num{1.17} &
	\num{1.21}$\pm$\num{0.05} &
	\num{1.86}$\pm$\num{0.1} &
	\num{0.53}$\pm$\num{0.083} &
	\num{1.89}$\pm$\num{0.13} &
	$\num{4.5}^{+\num{2.5}}_{-\num{0.45}}$
\\
	HATS-14 &
	\num{2.77}$\pm$\num{2.7e-06} &
	\num{12}$\pm$\num{3.9} &
	\num{0.97}$\pm$\num{0.02} &
	$\num{0.93}^{+\num{0.02}}_{-\num{0.01}}$ &
	\num{1.07}$\pm$\num{0.07} &
	$\num{1.04}^{+\num{0.032}}_{-\num{0.022}}$ &
	$>$\num{4.21}
\\
	HATS-15 &
	\num{1.75}$\pm$\num{9.4e-07} &
	\num{11}$\pm$\num{1.4} &
	\num{0.87}$\pm$\num{0.02} &
	\num{0.92}$\pm$\num{0.03} &
	\num{2.17}$\pm$\num{0.15} &
	\num{1.1}$\pm$\num{0.04} &
	$\num{6.28}^{+\num{0.237}}_{-\num{0.225}}$
\\
	HATS-16 &
	\num{2.69}$\pm$\num{1.1e-05} &
	\num{6.1}$\pm$\num{1} &
	\num{0.97}$\pm$\num{0.04} &
	$\num{1.24}^{+\num{0.1}}_{-\num{0.13}}$ &
	\num{3.27}$\pm$\num{0.19} &
	\num{1.3}$\pm$\num{0.15} &
	$\num{5.42}^{+\num{0.315}}_{-\num{0.334}}$
\\
	HATS-18 &
	\num{0.838}$\pm$\num{4.7e-07} &
	\num{9.4}$\pm$\num{0.5} &
	\num{1.04}$\pm$\num{0.047} &
	$\num{1.02}^{+\num{0.031}}_{-\num{0.057}}$ &
	\num{1.98}$\pm$\num{0.077} &
	$\num{1.34}^{+\num{0.049}}_{-\num{0.102}}$ &
	$\num{7.18}^{+\num{0.205}}_{-\num{0.173}}$
\\
	HATS-2 &
	\num{1.35}$\pm$\num{1e-06} &
	\num{12.5}$\pm$\num{0.5} &
	\num{0.88}$\pm$\num{0.04} &
	\num{0.9}$\pm$\num{0.02} &
	\num{1.3}$\pm$\num{0.15} &
	\num{0.776}$\pm$\num{0.055} &
	$\num{6.36}^{+\num{0.417}}_{-\num{0.234}}$
\\
	HATS-23 &
	\num{2.16}$\pm$\num{4.5e-06} &
	\num{13}$\pm$\num{1.6} &
	\num{1.12}$\pm$\num{0.05} &
	$\num{1.2}^{+\num{0.06}}_{-\num{0.08}}$ &
	\num{1.47}$\pm$\num{0.072} &
	$\num{1.9}^{+\num{0.3}}_{-\num{0.4}}$ &
	$\num{6.04}^{+\num{0.426}}_{-\num{0.427}}$
\\
	HATS-30 &
	\num{3.17}$\pm$\num{2.6e-06} &
	\num{13}$\pm$\num{1.7} &
	\num{1.09}$\pm$\num{0.03} &
	\num{1.06}$\pm$\num{0.04} &
	\num{0.706}$\pm$\num{0.039} &
	\num{1.18}$\pm$\num{0.052} &
	$>$\num{4.10}
\\
	HATS-32 &
	\num{2.81}$\pm$\num{5.5e-06} &
	\num{16}$\pm$\num{3.2} &
	\num{1.1}$\pm$\num{0.04} &
	$\num{1.1}^{+\num{0.1}}_{-\num{0.06}}$ &
	\num{0.92}$\pm$\num{0.1} &
	$\num{1.25}^{+\num{0.144}}_{-\num{0.096}}$ &
	$>$\num{4.88}
\\
	HATS-33 &
	\num{2.55}$\pm$\num{6.1e-06} &
	\num{19}$\pm$\num{1} &
	\num{1.06}$\pm$\num{0.03} &
	$\num{1.02}^{+\num{0.05}}_{-\num{0.04}}$ &
	\num{1.19}$\pm$\num{0.053} &
	$\num{1.23}^{+\num{0.112}}_{-\num{0.081}}$ &
	$>$\num{5.78}
\\
	HATS-34 &
	\num{2.11}$\pm$\num{4.7e-06} &
	\num{12}$\pm$\num{1.8} &
	\num{0.95}$\pm$\num{0.03} &
	\num{0.98}$\pm$\num{0.05} &
	\num{0.941}$\pm$\num{0.072} &
	\num{1.4}$\pm$\num{0.19} &
	$\num{5.4}^{+\num{1.9}}_{-\num{0.86}}$
\\
	HATS-4 &
	\num{2.52}$\pm$\num{2e-06} &
	\num{67}$\pm$\num{48} &
	\num{1}$\pm$\num{0.02} &
	\num{0.93}$\pm$\num{0.02} &
	\num{1.32}$\pm$\num{0.028} &
	\num{1.02}$\pm$\num{0.037} &
	$>$\num{6.24}
\\
	HATS-9 &
	\num{1.92}$\pm$\num{5.2e-06} &
	\num{17}$\pm$\num{3.3} &
	\num{1.03}$\pm$\num{0.039} &
	$\num{1.5}^{+\num{0.043}}_{-\num{0.101}}$ &
	\num{0.837}$\pm$\num{0.029} &
	\num{1.06}$\pm$\num{0.098} &
	$>$\num{4.33}
\\
	K2-60 &
	\num{3}$\pm$\num{4e-05} &
	\num{26}$\pm$\num{6} &
	\num{0.97}$\pm$\num{0.07} &
	\num{1.12}$\pm$\num{0.05} &
	\num{0.426}$\pm$\num{0.037} &
	\num{0.683}$\pm$\num{0.037} &
	$>$\num{4.58}
\\
	KELT-14 &
	$\num{1.71}^{+\num{3.2e-06}}_{-\num{2.6e-06}}$ &
	\num{23}$\pm$\num{5.7} &
	\num{1.24}$\pm$\num{0.04} &
	\num{1.52}$\pm$\num{0.03} &
	\num{1.28}$\pm$\num{0.032} &
	\num{1.74}$\pm$\num{0.047} &
	$>$\num{6.15}
\\
	KELT-8 &
	\num{3.24}$\pm$\num{0.00016} &
	\num{23}$\pm$\num{9.4} &
	$\num{1.21}^{+\num{0.08}}_{-\num{0.07}}$ &
	$\num{1.67}^{+\num{0.14}}_{-\num{0.12}}$ &
	$\num{0.867}^{+\num{0.065}}_{-\num{0.061}}$ &
	$\num{1.86}^{+\num{0.18}}_{-\num{0.16}}$ &
	$>$\num{4.17}
\\
	Kepler-17 &
	\num{1.49}$\pm$\num{2e-07} &
	\num{12.2}$\pm$\num{0.029} &
	\num{1.16}$\pm$\num{0.06} &
	\num{1.05}$\pm$\num{0.03} &
	\num{2.45}$\pm$\num{0.11} &
	\num{1.31}$\pm$\num{0.02} &
	$\num{7.14}^{+\num{0.466}}_{-\num{0.363}}$
\\
	Kepler-41 &
	\num{1.86}$\pm$\num{5.2e-07} &
	\num{11}$\pm$\num{3.6} &
	\num{1.15}$\pm$\num{0.04} &
	\num{1.29}$\pm$\num{0.02} &
	\num{0.56}$\pm$\num{0.08} &
	\num{1.29}$\pm$\num{0.02} &
	$>$\num{4.42}
\\
	Kepler-423 &
	\num{2.68}$\pm$\num{7e-08} &
	\num{22}$\pm$\num{0.121} &
	\num{0.85}$\pm$\num{0.04} &
	\num{0.95}$\pm$\num{0.04} &
	\num{0.59}$\pm$\num{0.081} &
	\num{1.19}$\pm$\num{0.052} &
	$\num{5.3}^{+\num{0.189}}_{-\num{0.21}}$
\\
	Kepler-44 &
	\num{3.25}$\pm$\num{3e-06} &
	\num{17}$\pm$\num{8.6} &
	\num{1.12}$\pm$\num{0.08} &
	\num{1.35}$\pm$\num{0.08} &
	\num{1}$\pm$\num{0.1} &
	\num{1.09}$\pm$\num{0.07} &
	$>$\num{4.46}
\\
	Kepler-45 &
	\num{2.46}$\pm$\num{4e-06} &
	\num{15.8}$\pm$\num{0.002} &
	\num{0.59}$\pm$\num{0.06} &
	\num{0.55}$\pm$\num{0.11} &
	\num{0.51}$\pm$\num{0.09} &
	\num{0.96}$\pm$\num{0.11} &
	$>$\num{4.19}
\\
	Qatar-1 &
	\num{1.42}$\pm$\num{2.2e-07} &
	\num{23.7}$\pm$\num{0.5} &
	\num{0.84}$\pm$\num{0.04} &
	\num{0.8}$\pm$\num{0.02} &
	$\num{1.29}^{+\num{0.052}}_{-\num{0.049}}$ &
	$\num{1.14}^{+\num{0.026}}_{-\num{0.025}}$ &
	$\num{7.21}^{+\num{0.639}}_{-\num{0.293}}$
\\
	Qatar-2 &
	\num{1.34}$\pm$\num{2.6e-07} &
	\num{11.4}$\pm$\num{0.5} &
	\num{0.74}$\pm$\num{0.02} &
	\num{0.78}$\pm$\num{0.01} &
	\num{2.49}$\pm$\num{0.054} &
	\num{1.25}$\pm$\num{0.013} &
	$\num{6.78}^{+\num{0.0725}}_{-\num{0.0729}}$
\\
	TrES-2 &
	\num{2.47}$\pm$\num{9e-08} &
	\num{25}$\pm$\num{19} &
	\num{0.98}$\pm$\num{0.06} &
	$\num{1}^{+\num{0.04}}_{-\num{0.03}}$ &
	\num{1.2}$\pm$\num{0.068} &
	\num{1.22}$\pm$\num{0.041} &
	$>$\num{5.10}
\\
	TrES-3 &
	\num{1.30618581} &
	\num{28}$\pm$\num{19} &
	\num{0.93}$\pm$\num{0.05} &
	\num{0.83}$\pm$\num{0.02} &
	$\num{1.91}^{+\num{0.075}}_{-\num{0.08}}$ &
	$\num{1.34}^{+\num{0.031}}_{-\num{0.037}}$ &
	$>$\num{6.61}
\\
	TrES-5 &
	\num{1.48}$\pm$\num{6.14e-06} &
	\num{11.6}$\pm$\num{1.11} &
	\num{0.9}$\pm$\num{0.03} &
	\num{0.87}$\pm$\num{0.01} &
	\num{1.79}$\pm$\num{0.068} &
	\num{1.21}$\pm$\num{0.021} &
	$\num{6.37}^{+\num{0.263}}_{-\num{0.251}}$
\\
	WASP-104 &
	$\num{1.76}^{+\num{1.8e-06}}_{-\num{3.6e-06}}$ &
	\num{1e+02}$\pm$\num{2e+02} &
	\num{1.08}$\pm$\num{0.05} &
	\num{0.96}$\pm$\num{0.03} &
	\num{1.27}$\pm$\num{0.047} &
	\num{1.14}$\pm$\num{0.037} &
	$>$\num{6.76}
\\
	WASP-114 &
	$\num{1.55}^{+\num{1.2e-06}}_{-\num{9.1e-07}}$ &
	\num{11}$\pm$\num{1.3} &
	\num{1.29}$\pm$\num{0.05} &
	\num{1.43}$\pm$\num{0.06} &
	\num{1.77}$\pm$\num{0.064} &
	\num{1.34}$\pm$\num{0.064} &
	$\num{6.41}^{+\num{0.319}}_{-\num{0.293}}$
\\
	WASP-119 &
	\num{2.5}$\pm$\num{1e-05} &
	\num{9e+01}$\pm$\num{1e+02} &
	\num{1.02}$\pm$\num{0.06} &
	\num{1.2}$\pm$\num{0.1} &
	\num{1.23}$\pm$\num{0.08} &
	\num{1.4}$\pm$\num{0.2} &
	$>$\num{6.22}
\\
	WASP-123 &
	\num{2.98}$\pm$\num{2.3e-06} &
	\num{65}$\pm$\num{45} &
	\num{1.17}$\pm$\num{0.06} &
	\num{1.28}$\pm$\num{0.05} &
	\num{0.899}$\pm$\num{0.036} &
	\num{1.33}$\pm$\num{0.074} &
	$>$\num{5.92}
\\
	WASP-124 &
	\num{3.37}$\pm$\num{1e-06} &
	\num{16}$\pm$\num{4.5} &
	\num{1.07}$\pm$\num{0.05} &
	\num{1.02}$\pm$\num{0.02} &
	\num{0.6}$\pm$\num{0.07} &
	\num{1.24}$\pm$\num{0.03} &
	$>$\num{4.16}
\\
	WASP-133 &
	\num{2.18}$\pm$\num{1e-06} &
	\num{2e+02}$\pm$\num{6e+02} &
	\num{1.16}$\pm$\num{0.08} &
	\num{1.44}$\pm$\num{0.05} &
	\num{1.16}$\pm$\num{0.09} &
	\num{1.21}$\pm$\num{0.05} &
	$>$\num{7.02}
\\
	WASP-135 &
	\num{1.4}$\pm$\num{8e-07} &
	\num{10}$\pm$\num{2.1} &
	\num{0.98}$\pm$\num{0.06} &
	\num{0.96}$\pm$\num{0.05} &
	\num{1.9}$\pm$\num{0.08} &
	\num{1.3}$\pm$\num{0.09} &
	$\num{6.57}^{+\num{0.541}}_{-\num{0.557}}$
\\
	WASP-140 &
	\num{2.24}$\pm$\num{8e-07} &
	\num{10.4}$\pm$\num{0.5} &
	\num{0.9}$\pm$\num{0.04} &
	\num{0.87}$\pm$\num{0.04} &
	\num{2.44}$\pm$\num{0.07} &
	$\num{1.4}^{+\num{0.42}}_{-\num{0.18}}$ &
	$\num{5.82}^{+\num{0.181}}_{-\num{0.149}}$
\\
	WASP-141 &
	\num{3.31}$\pm$\num{5e-06} &
	\num{18}$\pm$\num{3.8} &
	\num{1.25}$\pm$\num{0.06} &
	\num{1.37}$\pm$\num{0.07} &
	\num{2.69}$\pm$\num{0.15} &
	\num{1.21}$\pm$\num{0.08} &
	$>$\num{5.86}
\\
	WASP-19 &
	\num{0.789}$\pm$\num{4e-08} &
	\num{11.8}$\pm$\num{0.5} &
	\num{0.9}$\pm$\num{0.04} &
	\num{1}$\pm$\num{0.02} &
	$\num{1.07}^{+\num{0.038}}_{-\num{0.037}}$ &
	\num{1.39}$\pm$\num{0.032} &
	$\num{6.86}^{+\num{0.0491}}_{-\num{0.0472}}$
\\
	WASP-23 &
	$\num{2.94}^{+\num{1.1e-06}}_{-\num{1.3e-06}}$ &
	\num{18}$\pm$\num{2.7} &
	\num{0.78}$\pm$\num{0.13} &
	\num{0.77}$\pm$\num{0.05} &
	$\num{0.884}^{+\num{0.088}}_{-\num{0.099}}$ &
	$\num{0.962}^{+\num{0.047}}_{-\num{0.056}}$ &
	$\num{6}^{+\num{1.6}}_{-\num{1.4}}$
\\
	WASP-26 &
	\num{2.76}$\pm$\num{6.7e-06} &
	\num{17}$\pm$\num{1.9} &
	\num{1.11}$\pm$\num{0.03} &
	\num{1.3}$\pm$\num{0.06} &
	\num{1.03}$\pm$\num{0.021} &
	\num{1.28}$\pm$\num{0.075} &
	$\num{5.64}^{+\num{0.34}}_{-\num{0.298}}$
\\
	WASP-36 &
	\num{1.54}$\pm$\num{2.4e-07} &
	\num{15}$\pm$\num{5.5} &
	\num{1.08}$\pm$\num{0.03} &
	\num{0.98}$\pm$\num{0.01} &
	\num{2.36}$\pm$\num{0.07} &
	\num{1.27}$\pm$\num{0.03} &
	$>$\num{5.99}
\\
	WASP-4 &
	\num{1.34}$\pm$\num{3.1e-07} &
	\num{22.2}$\pm$\num{0.5} &
	\num{0.89}$\pm$\num{0.01} &
	\num{0.92}$\pm$\num{0.06} &
	\num{1.22}$\pm$\num{0.013} &
	\num{1.21}$\pm$\num{0.07} &
	$\num{7.09}^{+\num{0.262}}_{-\num{0.204}}$
\\
	WASP-41 &
	\num{3.05}$\pm$\num{9e-07} &
	\num{18.4}$\pm$\num{0.5} &
	\num{0.93}$\pm$\num{0.07} &
	\num{0.87}$\pm$\num{0.03} &
	\num{0.94}$\pm$\num{0.05} &
	\num{1.21}$\pm$\num{0.07} &
	$\num{5.31}^{+\num{1.27}}_{-\num{0.387}}$
\\
	WASP-43 &
	\num{0.813}$\pm$\num{1e-06} &
	\num{15.6}$\pm$\num{0.5} &
	\num{0.58}$\pm$\num{0.05} &
	\num{0.6}$\pm$\num{0.04} &
	\num{1.78}$\pm$\num{0.1} &
	$\num{0.93}^{+\num{0.07}}_{-\num{0.09}}$ &
	$\num{7.6}^{+\num{0.185}}_{-\num{0.189}}$
\\
	WASP-44 &
	\num{2.42}$\pm$\num{8.7e-06} &
	\num{15}$\pm$\num{4.3} &
	\num{0.95}$\pm$\num{0.03} &
	\num{0.93}$\pm$\num{0.07} &
	\num{0.889}$\pm$\num{0.062} &
	$\num{1.1}^{+\num{0.13}}_{-\num{0.14}}$ &
	$>$\num{4.26}
\\
	WASP-46 &
	\num{1.43}$\pm$\num{9.3e-07} &
	\num{16.1}$\pm$\num{1} &
	\num{0.83}$\pm$\num{0.08} &
	\num{0.86}$\pm$\num{0.03} &
	\num{1.91}$\pm$\num{0.13} &
	\num{1.33}$\pm$\num{0.058} &
	$\num{7}^{+\num{0.252}}_{-\num{0.247}}$
\\
	WASP-49 &
	\num{2.78}$\pm$\num{5.6e-06} &
	\num{55}$\pm$\num{18} &
	\num{0.94}$\pm$\num{0.08} &
	\num{0.98}$\pm$\num{0.03} &
	\num{0.378}$\pm$\num{0.027} &
	\num{1.11}$\pm$\num{0.047} &
	$>$\num{5.69}
\\
	WASP-5 &
	$\num{1.63}^{+\num{2.2e-06}}_{-\num{4.9e-06}}$ &
	\num{17}$\pm$\num{2.5} &
	$\num{0.96}^{+\num{0.13}}_{-\num{0.09}}$ &
	$\num{1.03}^{+\num{0.06}}_{-\num{0.07}}$ &
	$\num{1.58}^{+\num{0.13}}_{-\num{0.1}}$ &
	\num{1.15}$\pm$\num{0.048} &
	$\num{6.22}^{+\num{0.33}}_{-\num{0.329}}$
\\
	WASP-50 &
	\num{1.96}$\pm$\num{5.1e-06} &
	\num{16.3}$\pm$\num{0.5} &
	\num{0.89}$\pm$\num{0.08} &
	\num{0.84}$\pm$\num{0.03} &
	$\num{1.47}^{+\num{0.091}}_{-\num{0.086}}$ &
	\num{1.15}$\pm$\num{0.048} &
	$\num{6.08}^{+\num{0.113}}_{-\num{0.12}}$
\\
	WASP-52 &
	\num{1.75}$\pm$\num{1.2e-06} &
	\num{16.4}$\pm$\num{0.5} &
	\num{0.87}$\pm$\num{0.03} &
	\num{0.79}$\pm$\num{0.02} &
	\num{0.46}$\pm$\num{0.02} &
	\num{1.27}$\pm$\num{0.03} &
	$>$\num{4.54}
\\
	WASP-57 &
	\num{2.84}$\pm$\num{8.1e-07} &
	\num{13}$\pm$\num{4.5} &
	\num{0.89}$\pm$\num{0.07} &
	\num{0.93}$\pm$\num{0.03} &
	\num{0.644}$\pm$\num{0.062} &
	$\num{0.916}^{+\num{0.017}}_{-\num{0.014}}$ &
	$>$\num{3.30}
\\
	WASP-6 &
	\num{3.36}$\pm$\num{3.1e-07} &
	\num{23.8}$\pm$\num{0.5} &
	\num{0.84}$\pm$\num{0.06} &
	\num{0.86}$\pm$\num{0.02} &
	\num{0.485}$\pm$\num{0.027} &
	\num{0.86}$\pm$\num{0.12} &
	$>$\num{4.93}
\\
	WASP-64 &
	\num{1.57}$\pm$\num{1.5e-06} &
	\num{16}$\pm$\num{3.7} &
	\num{1}$\pm$\num{0.03} &
	\num{1.06}$\pm$\num{0.03} &
	\num{1.27}$\pm$\num{0.068} &
	\num{1.27}$\pm$\num{0.039} &
	$\num{6.37}^{+\num{0.536}}_{-\num{0.564}}$
\\
	WASP-65 &
	\num{2.31}$\pm$\num{1.5e-06} &
	\num{14}$\pm$\num{2.1} &
	\num{0.93}$\pm$\num{0.14} &
	\num{1.01}$\pm$\num{0.05} &
	\num{1.6}$\pm$\num{0.16} &
	\num{1.11}$\pm$\num{0.059} &
	$\num{5.91}^{+\num{0.244}}_{-\num{0.252}}$
\\
	WASP-77 A &
	\num{1.36}$\pm$\num{2e-06} &
	\num{15.4}$\pm$\num{0.5} &
	\num{1}$\pm$\num{0.04} &
	\num{0.95}$\pm$\num{0.01} &
	\num{1.76}$\pm$\num{0.06} &
	\num{1.21}$\pm$\num{0.02} &
	$\num{6.87}^{+\num{0.0745}}_{-\num{0.0763}}$
\\
	WASP-80 &
	$\num{3.07}^{+\num{8.3e-07}}_{-\num{7.9e-07}}$ &
	\num{24}$\pm$\num{3} &
	\num{0.58}$\pm$\num{0.05} &
	\num{0.59}$\pm$\num{0.02} &
	$\num{0.538}^{+\num{0.035}}_{-\num{0.036}}$ &
	$\num{0.952}^{+\num{0.026}}_{-\num{0.027}}$ &
	$\num{5.4}^{+\num{3.1}}_{-\num{1.3}}$
\\
	WASP-81 &
	\num{2.72}$\pm$\num{2.3e-06} &
	\num{54}$\pm$\num{33} &
	\num{1.08}$\pm$\num{0.06} &
	\num{1.28}$\pm$\num{0.04} &
	$\num{0.729}^{+\num{0.036}}_{-\num{0.035}}$ &
	$\num{1.43}^{+\num{0.051}}_{-\num{0.046}}$ &
	$>$\num{5.72}
\\
	WASP-95 &
	\num{2.18}$\pm$\num{1.4e-06} &
	\num{20.7}$\pm$\num{2} &
	\num{1.11}$\pm$\num{0.09} &
	$\num{1.13}^{+\num{0.08}}_{-\num{0.04}}$ &
	$\num{1.13}^{+\num{0.1}}_{-\num{0.04}}$ &
	\num{1.21}$\pm$\num{0.06} &
	$>$\num{5.74}
\\
	WASP-96 &
	\num{3.43}$\pm$\num{2.7e-06} &
	\num{35}$\pm$\num{31} &
	\num{1.06}$\pm$\num{0.09} &
	\num{1.05}$\pm$\num{0.05} &
	\num{0.48}$\pm$\num{0.03} &
	\num{1.2}$\pm$\num{0.06} &
	$>$\num{4.28}
\\
	WASP-97 &
	\num{2.07}$\pm$\num{1e-06} &
	\num{49}$\pm$\num{22} &
	\num{1.12}$\pm$\num{0.06} &
	\num{1.06}$\pm$\num{0.04} &
	\num{1.32}$\pm$\num{0.05} &
	\num{1.13}$\pm$\num{0.06} &
	$>$\num{6.49}
\\
	WASP-98 &
	\num{2.96}$\pm$\num{4.3e-07} &
	\num{4e+01}$\pm$\num{4e+01} &
	\num{0.81}$\pm$\num{0.06} &
	\num{0.74}$\pm$\num{0.02} &
	\num{0.922}$\pm$\num{0.08} &
	\num{1.1}$\pm$\num{0.04} &
	$\num{6.5}^{+\num{1.8}}_{-\num{1.5}}$
\\
	WTS-2 &
	\num{1.02}$\pm$\num{6.5e-07} &
	\num{17}$\pm$\num{7.9} &
	\num{0.82}$\pm$\num{0.08} &
	\num{0.75}$\pm$\num{0.03} &
	\num{1.1}$\pm$\num{0.16} &
	\num{1.36}$\pm$\num{0.061} &
	$>$\num{5.84}
\\
	XO-2 N &
	\num{2.62}$\pm$\num{2.8e-07} &
	\num{48}$\pm$\num{5.7} &
	\num{0.97}$\pm$\num{0.05} &
	$\num{1.01}^{+\num{0.1}}_{-\num{0.07}}$ &
	\num{0.597}$\pm$\num{0.021} &
	\num{1.02}$\pm$\num{0.031} &
	$>$\num{6.11}
\\
\enddata
\end{deluxetable}

\bibliography{bibliography}{}
\bibliographystyle{aasjournal}

\end{document}